\documentclass[prl,reprint,groupedaddress,showkeys]{revtex4-1}
\usepackage{graphicx}
\usepackage{enumerate}
\usepackage{array}
\usepackage{amssymb,amsfonts,amsmath}
\usepackage{subfig}
\usepackage{dcolumn}
\usepackage{bm}

\begin{document}

%----------------------------------------------------------------------------------------
%	TITLE AND AUTHORS
%----------------------------------------------------------------------------------------

\title{Network cluster detecting in associated bi-graph view} % For titles, only capitalize the first letter

%------------------------------------------------
\author{Zhe He}
\email{2287580716@qq.com, ychezhe@mail.ustc.edu.cn}
\author{Yi-Ming Huang}
\author{Rui-Jie Xu}
\author{Bing-Hong Wang}
\email{bhwang@ustc.edu.cn}
\altaffiliation{\\School of Physics, University of Science and Technology of China,Hefei 230026, People's Republic of China}
\author{Zhong-Can Ou-Yang}
\email{oy@itp.ac.cn}
\altaffiliation{\\Institute for Theoretical Physics, Chinese Academy of Sciences,Beijing 100080, People's Republic of China}
%% Enter authors via the \author command.
%% Use \affil to define affiliations.
%% (Leave no spaces between author name and \affil command)

%% Note that the \thanks{} command has been disabled in favor of
%% a generic, reserved space for PNAS publication footnotes.

%% \author{<author name>
%% \affil{<number>}{<Institution>}} One number for each institution.
%% The same number should be used for authors that
%% are affiliated with the same institution, after the first time
%% only the number is needed, ie, \affil{number}{text}, \affil{number}{}
%% Then, before last author ...
%% \and
%% \author{<author name>
%% \affil{<number>}{}}

%% For example, assuming Garcia and Sonnery are both affiliated with
%% Universidad de Murcia:
%% \author{Roberta Graff\affil{1}{University of Cambridge, Cambridge,
%% United Kingdom},
%% Javier de Ruiz Garcia\affil{2}{Universidad de Murcia, Bioquimica y Biologia
%% Molecular, Murcia, Spain}, \and Franklin Sonnery\affil{2}{}}

%----------------------------------------------------------------------------------------

 % The \maketitle command is necessary to build the title page
\thanks{\\Thank my fellow Jia-Rong Xie and thank Prf. Zeng-Ru Di from Beijing Normal University. This work is funded by the National Natural Science Foundation of China (Grant Nos. : 11275186, 91024026, FOM2014OF001 )}
%\maketitle
%\begin{article}

%----------------------------------------------------------------------------------------
%	ABSTRACT, KEYWORDS AND ABBREVIATIONS
%----------------------------------------------------------------------------------------

\begin{abstract}
We find there is relationship between the associated bigraph and the cluster (or community) detecting on network. By imbedding the associated bigraph of some network (suppose it has cluster structures) into some space, we can identify the clusters on this network ,which is a new method for network cluster detecting. And this method, of which the physical meaning is clear and the time complexity is acceptable, may provide us a new point to understand the structure and character of networks. In this paper, We test the methods on serval computer-generated networks and real networks. A computer-generated network with 128 vertexes and the Zachary Network, which presents the structure of a karate club, can be partitioned correctly by these methods. And the Dolphin network, which presents the relationship between 62 dolphins on the coast of New Zealand, is partitioned reasonably.
\end{abstract}

%------------------------------------------------

\keywords{cluster, community, networks, associated bigraph} % When adding keywords, separate each term with a straight line: |

%------------------------------------------------

%% Optional for entering abbreviations, separate the abbreviation from
%% its definition with a comma, separate each pair with a semicolon:
%% for example:
%% \abbreviations{SAM, self-assembled monolayer; OTS,
%% octadecyltrichlorosilane}

% \abbreviations{}
%\abbreviations{SAM, self-assembled monolayer; OTS, octadecyltrichlorosilane}

%----------------------------------------------------------------------------------------
%	PUBLICATION CONTENT
%----------------------------------------------------------------------------------------

%% The first letter of the article should be drop cap: \dropcap{} e.g.,
%\dropcap{I}n this article we study the evolution of ''almost-sharp'' fronts
\maketitle
\section{Introduction}

As a essential problem in network science, network cluster detecting is significant for computer science\cite{1}, biology\cite{2,7,24}, communication and social networks\cite{3,4} and marketing strategy\cite{5}. And it gains lots of attention from researchers in related fields. Especially in recent years, understanding networks deeper and deeper, people get rich harvests in the study of network cluster detecting\cite{6}. Many detecting algorithms and evaluation criteria is proposed. Some of the algorithms are based on some operating process on network structure\cite{7,8,9,23}, some are based on spectrum analysis algorithm\cite{10,11} and some are based on network dynamics\cite{6,12,24} and so on. The criteria various from Q function\cite{13} to association quality, overlapping quality\cite{14} and Benchmark graphs\cite{25}, etc. From the angle of category, there are overlapping clustering\cite{15} and non-overlapping clustering\cite{7,8,9}.
\\\indent
The new network cluster detecting method we put forward is based on a measure on associated bigraph(AG). In this paper, discrete equidistant imbedding(DEI) and continuous imbedding(CI) separately provide two different measures and two different methods.

%------------------------------------------------

\section{AG and THE TWO METHODS}
In this section, we give a definition to Associated Bigraph(AG) and from it we propose two methods, DEI method and CI method. And to test the methods, we give partitions to a computer-generated network, Zachary network\cite{17} and Dolphin network\cite{18}. At last, we compare the results by our methods with those by modularity method in Gephi\cite{22}.
\subsection{Definition of AG and DEI}
\begin{figure}[!ht]\centering
  \includegraphics[scale=0.22]{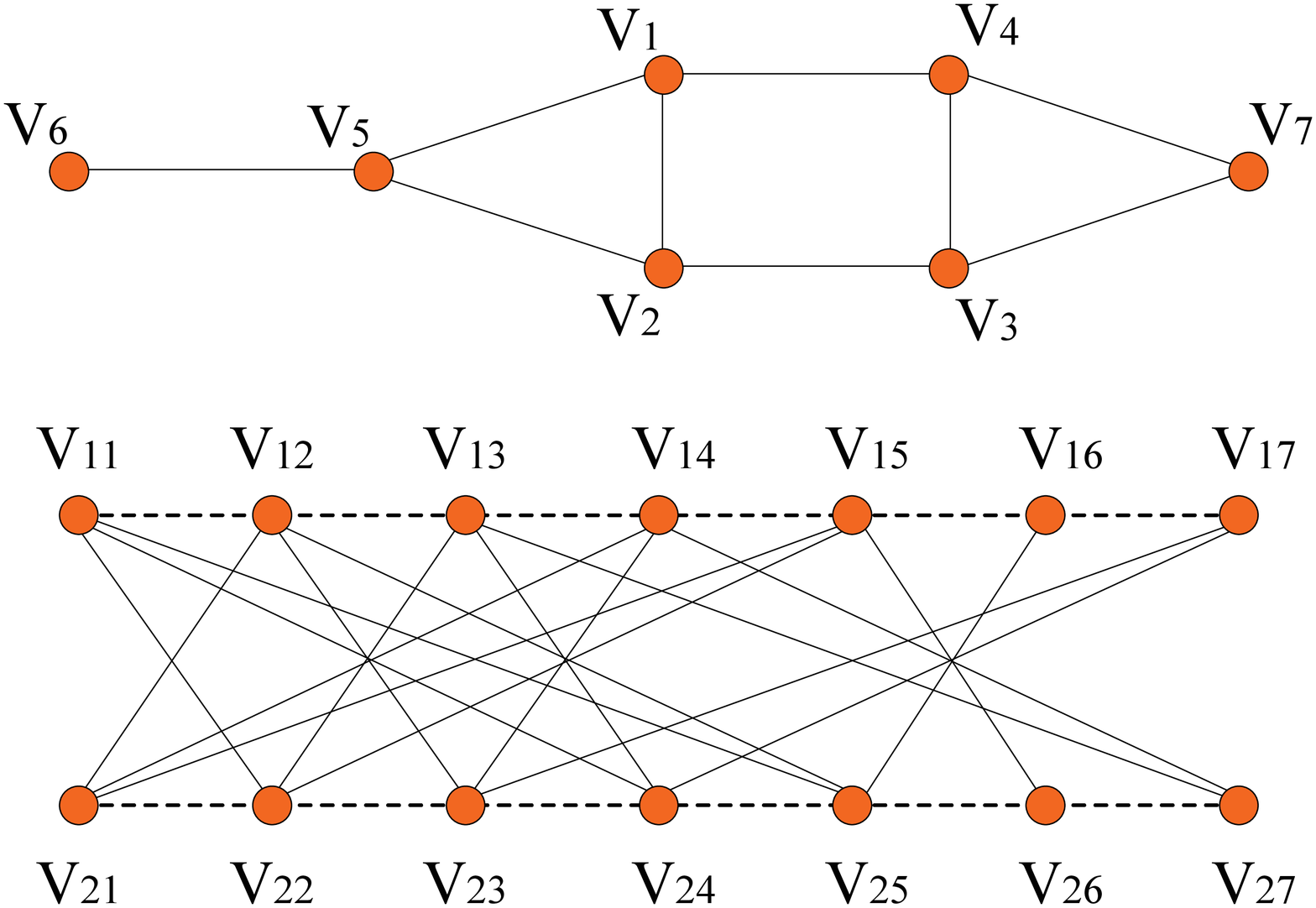}
  \caption{Graph and its AG are equidistantly imbedded on the two lines.}\label{fg:1}
\end{figure}

Suppose graph $G=(V,E)$, $V$ refers to its vertex set, $E$ refers to its edge set, $|V|=N$.
Then, the associated bigraph of $G$ is $G_A=(V_1\bigcup V_2, E_A)$. If $V=\{v_1,v_2,\cdots,v_N\}$, ~then $V_1=\{v_{11},v_{12},\cdots,v_{1N}\}$,~$V_2=\{v_{21},v_{22},\cdots,v_{2N}\}$,~for any $i$, $v_i$ corresponds to $v_{1i}$ and $v_{2i}$. If and only if $(v_i,v_j)\in E$,~$(v_{1i},v_{2j})\in E_A$. It is easy to know, $G_A$ is a bigraph. $V_1$ and $V_2$ are two parts of it, $|E_A|=|E|$(suppose undirected edges be bidirectional edges). If we merge the corresponding vertexes in $V_1$ and $V_2$, $G_A$ is equal to $G$. See FIG.\ref{fg:1}.
We place the vertexes of AG as in FIG.\ref{fg:1}. With equal interval, place the vertexes from sets $V_1$, $V_2$ on two parallel lines $(L_1\&L_2)$ and let those with corresponding labels be at corresponding positions. We call the placing pattern described above discrete equidistant imbedding (DEI). Of a given graph, the AG has $N!/2$ kinds of different DEIs.( If the graph has some symmetry, the number will decrease.) Without loss of generality, we let the allowed coordinate of vertexes in DEI successively be $1,2,3\dots N$. Thus, the distance between adjacent vertexes is $1$.
\subsection{DEI method}
Now, we consider the simple graphs(undirected, non-weighted, acyclic, non-multiple edges). If there are clusters structures in such graphs, among different DEIs of an AG, at least there is one that the vertexes are arrayed in the sequence of cluster. That is to say, vertexes of a same cluster will be placed together. In detail, different clusters will be placed nearer if they have closer relations. In the interior of a cluster, vertexes with closer relations are placed nearer. This arrangement is called optimal DEI.
\\\indent
We define the distance in DEI as follow: the distance between $v_{1i}$ and $v_{1j}$ is $|xi-xj|$, where $x_i$ and $x_j$ are the coordinates of $v_{1i}$ and $v_{2j}$. If edge $(v_{1i},v_{2j})$ exists, we define the length of $(v_{1i},v_{2j})$ as $|xi-xj|$. Let
\begin{equation}
Z = \sum\limits_{ij} {{a_{ij}}| {{x_i} - {x_j}}|}.
\label{eq:1}
\end{equation}

\begin{figure*}
\centering
\includegraphics[scale=0.42]{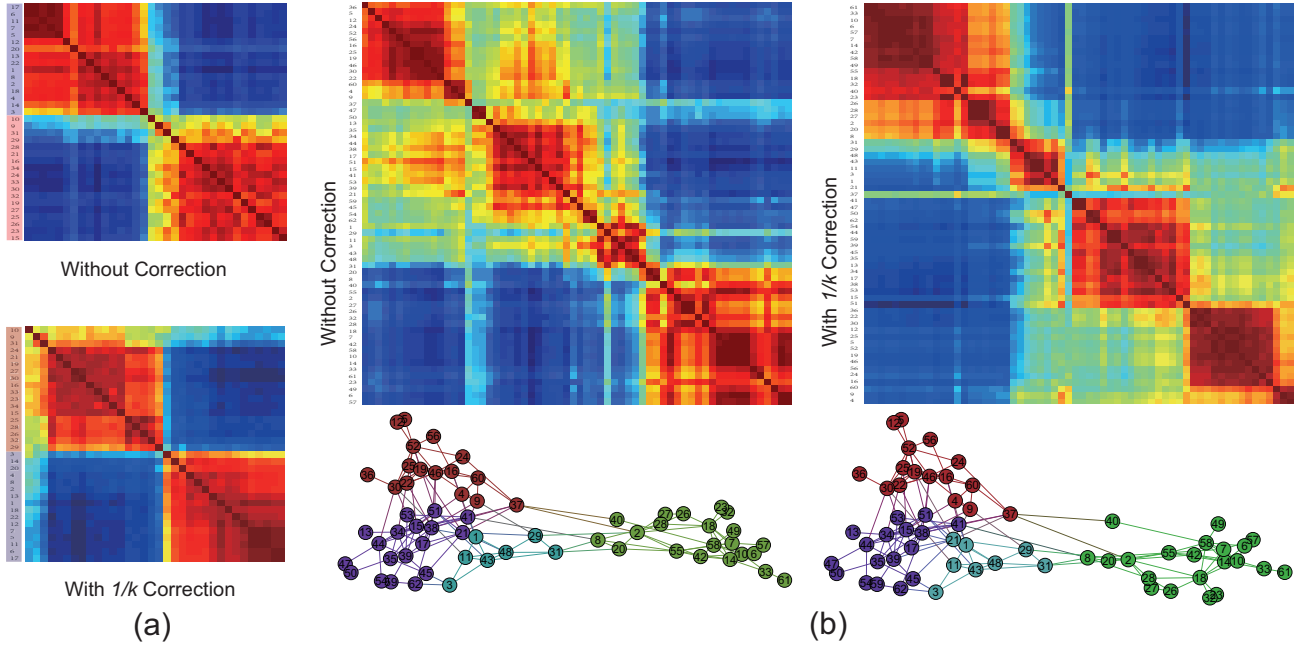}
\caption{(a) are graphs of R matrix of Zachary Network without correction and with 1/k correction. The left number refers to a vertex label. In the history, the club divided into two parts. R matrices can correctly distinguish them.
\\
(b) shows R matrice of Dolphin Network without correction and with 1/k correction and clusters labeled by colors.}\label{fg:2}
\end{figure*}

We treat Z as an objective function, and minimize it under the condition of DEI, the solution of which is the optimal DEI.
\\\indent
If an edge $a$ connects $v_{1i}$ and $v_{2j}$(suppose $i<j$), we can find it in the interval of $k$ and $k+1$(FIG.\ref{fg:1}), where $i<=k<k+1<=j$. The number of edges found in the interval of $k$ and $k+1$ is defined as the cross of $k$.
Let the number of crosses be $\{m_1,m_2,\cdots,m_{N-1}\}$, it is easy to see that $Z=\sum\limits_i {{m_i}}$, which means the optimal DEI corresponds to \textquoteleft the minimum sum of crosses\textquoteright.
\\\indent
This optimization equals to the follow operation in the adjacency matrix $A$: $|x_i-x_j|$ suggests the absolute difference of the element $a_{ij}$'s column number and row number, which can measure the distance of the element and main diagonal. In order to minimize $Z$, by swapping vertexes we move the non-zero elements to the main diagonal as near as possible.
\\\indent
Actually, this definition of $Z$ may give \textquoteleft greater rights\textquoteright ~to vertexes with larger degree. In order to minimize $Z$, some vertexes with large degree may draw connected vertexes close to themselves. This may drown the structures of other vertexes. Or say, edges from some vertexes is so large a proportion of total edges that these vertexes affect the arrangement too strong and the effect of other vertexes are unimportant. In order to avoid this, we have to correct $Z$. A natural correction is to average the weight of each vertex.
\begin{equation}
Z = \sum\limits_{ij} {\frac{{{a_{ij}}| {{x_i} - {x_j}}|}}{{{k_i}}}}.
\label{eq:2}
\end{equation}
Since it is a undirected graph, Eq(\ref{eq:2})is equivalent to
\begin{equation}
Z = \frac{1}{2}\sum\limits_{ij} {{a_{ij}}(\frac{1}{{{k_i}}} + \frac{1}{{{k_j}}})\left| {{x_i} - {x_j}} \right|}.
\label{eq:3}
\end{equation}
Although we correct $Z$ by $1/k$, people may have different opinions on whether this correction is reasonable or not. Some people may believe that vertex with large degree should have a greater effect. While, our suggestion is: the definition of clustering do not need to be unique. Different definitions should be allowed in different case. It is more important that a good definition should match the practical problem. However, the simulation results suggest that $Z$ corrected by $1/k$ has a higher resolution power. (FIG.\ref{fg:2})
\begin{figure}[!ht]
\centering
\includegraphics[scale=0.35]{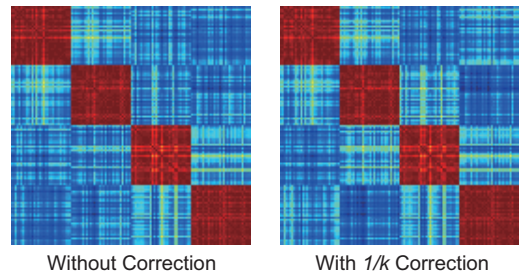}
\caption{are graphs of R matrix of computer-generating network without correction and with 1/k correction.
\\
 the network is generated as follow:
 First generate four ER networks of 32 vertexes with $p_1=0.6$. Then randomly construct edges among different networks with $p_2=0.2$\cite{40}. In this picture, vertexes are arranged in the order of four networks. We can see that our methods can uncover the four clusters correctly.}
\end{figure}

\begin{figure*}[!ht]\centering
\includegraphics[scale=0.54]{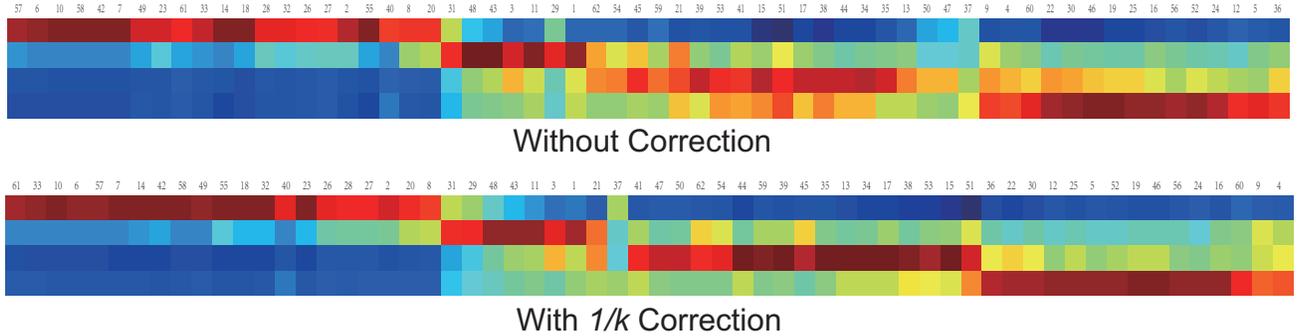}
\caption{We detect an overlapping vertex by comparing its average correlation coefficient with other vertexes in different clusters. The Dolphin Network is partitioned into four clusters. Vertex $21$, $51$ and $62$ are overlappings by the method without correction. Vertex $21$ and $51$ are overlappings by the method with $1/k$ correction.}\label{fg:3}
\end{figure*}
We use Simulated Annealing Algorithm to find the solution of minimum $Z$.
\\\indent
A consequent question is even though we have found the optimal DEI, how can we know which vertexes belong to the same cluster?
A intuitive idea is to count the crosses between adjacent vertexes. The sum of crosses between adjacent vertexes belonging to the same cluster will be larger than that between adjacent vertexes belonging to different clusters.(considering that a large degree vertex may drown the information of other vertexes, we can give each edge a weight $1/k$). However, after the simulation, we find that this method is not so effective and cannot show the \textquoteleft bond' of real cluster well. Thus, we adopt the following movement correlation method.
\\\indent
Suppose we move some vertex pair $(v_{1i},v_{2i})$ on $L_1$ and $L_2$, some other vertexes have to move in order to keep $Z$ as small as possible. If a vertex always follows another, it is that there is movement correlation between them. We use the strength of the movement correlation measured by Pearson correlation coefficient matrix R to partition clusters. Vertexes in the same cluster are with strong correlation.
where \[Rij = \frac{{Cov({x_i},{x_j})}}{{\sqrt {Cov({x_i},{x_i})} \sqrt {Cov({x_j},{x_j})} }}.\]
In the simulation, we randomly fix a small part of vertexes(e.g.5\%) and minimize $Z$. Repeat this operation several times, we can get many optimal DEIs in this case and calculate the coordinate correlation of different vertexes. Then we get matrix $R$.

It is worth mentioning that overlap is allowed in this method (FIG.\ref{fg:3}).
\\\indent
Although we get $R$, we do not have a clear-cut criterion for clustering. For example, in FIG.{\ref{fg:2}}, if we set different resolutions, we can get different results of clustering. Maybe two clusters, maybe three, maybe four. What partition is reasonable is worth of discussion. Further more, we have done an elementary analysis. We can consider this partition criterion from two aspects. 1.amplitude criterion, 2. step criterion. The first criterion means that we can set a value and when a element of matrix $R$ is less than this value, we set it to zero. At last a non-zero diagonal block is a cluster(allow overlapping). The second criterion means that we can consider the step (difference) of adjacent elements. We can confirm the \textquoteleft bond' of a cluster by finding a position with big step. We can achieve this by high-pass filtering. Criterion 2 is affected seriously by the sequence of vertexes in figure of matrix $R$. In fact, \textquoteleft how many the clusters is there' is just a question has more than one answer, for when the criterion is \textquoteleft loose', there may be two clusters , and when the criterion is \textquoteleft strict', there may be four clusters.
\\\indent
Actually, $Z$ in Eq.(\ref{eq:1}) and Eq.(\ref{eq:2}) is regarded as $L^1$ norm. We can generally define the measure based on $L^p$ norm. Here we write two possible definitions of $L^2$ norm.
\\
Uncorrected :
\[Z = \sum\limits_{ij} {{a_{ij}}(} {x_i} - {x_j}{)^2}.\]
Corrected:
\[Z = \sum\limits_{ij} {\frac{{{a_{ij}}{{({x_i} - {x_j})}^2}}}{{{k_i}}}}    or  Z = \sum\limits_{ij} {\frac{{{a_{ij}}{{({x_i} - {x_j})}^2}}}{{{k_i}^2}}} .\]
On the previous two real networks, similar partitions can be made by $L^2$ norm and by $L^1$ norm. But the resolution of $L^2$ norm method is lower than that of $L^1$ norm.
\subsection{CI method}
\begin{figure}[!ht]\centering
  \includegraphics[scale=0.28]{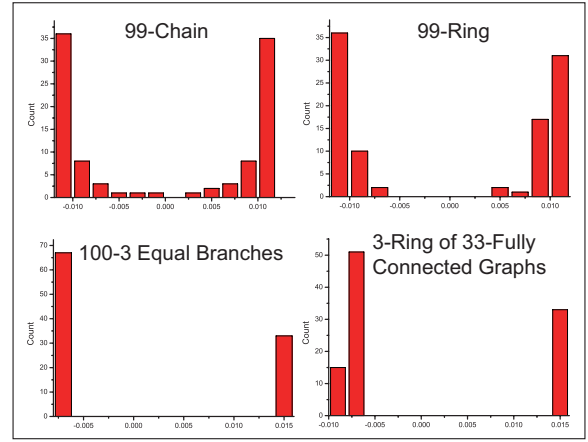}
  \caption{99-chain is a network of 99vertexes,and they link one by one like a chain. 99-ring is a ring of 99vertexes. The optimal CI divide 99-chain into two groups from the middle and divide 99-ring into two chains with equal amount of vertexes. 100-3 equal branches is a graph consisting of three 34-chains with a common vertex. The optimal CI randomly put two into a group. 3-Ring of 33 Fully Connected Graph is a 99 vertexes graph consisting of three fully Connected subgraphs of 33vertrexes. The three subgraphs link each other forming a ring with three symmetry. The optimal CI randomly put two subgraphs into one group.The interval of each bar is 0.002}\label{fg:4}
\end{figure}

We can change DEI to continuous imbedding(CI). CI means that a vertex can be placed at any point on the line and the vertex pair with same label still should has the same coordinate. Comparing CI and DEI, the objective function $Z$ does not change, but the feasible region changes from all arrangements of $\{1,2,\dots,N\}$ to $R^N$.
Corresponding to the discrete imbedding, here we set the constraints of continuous imbedding:
\\
For $L^1$ norm:
$1.\sum\limits_i {{x_i}}  = 0$,
$2.\sum\limits_i {\left| {{x_i}} \right|}  = 1$.\\
For $L^2$ norm:
$1.\sum\limits_i {{x_i}}  = 0$,
$2.{\sum\limits_i {{x_i}} ^2} = 1$.
\\
There is a special relation between $L^2$ norm CI method and spectral method\cite{10,11}.
Next, we only discuss $L^1$ case with $1/k$ correction.
\\\indent
For $L^1$ case with $1/k$ correction,in simulations, the vertexes with minimum $Z$ are always scattered in two groups (FIG.{\ref{fg:4}}).
Thus, we can put forward a CI method measured by $L^1$ norm with $1/k$ correction. The algorithm for a given network G is as follow:
\begin{description}
 \item[Step A] minimize Z of G and get the optimal solution X, of which the positive components the induced subgraph affiliating with is called G1,and the induced subgraph of remaining vertexes is called G2.
 \item[Step B] respectively redo Step A on G1 and G2 until each vertex is a induced subgraph.
\end{description}
This process generates a binary tree called cluster tree.
\\\indent
How can we take advantage of the cluster tree to uncover the clusters?\\\indent
Criteria are needed. Here, we adopt $Q$ function. In detail, we partition the leaves of the cluster tree with maximized $Q$ in all possible ways and each part is a cluster.
\begin{figure}[!ht]
\centering
  \subfloat[]{\includegraphics[scale=0.3]{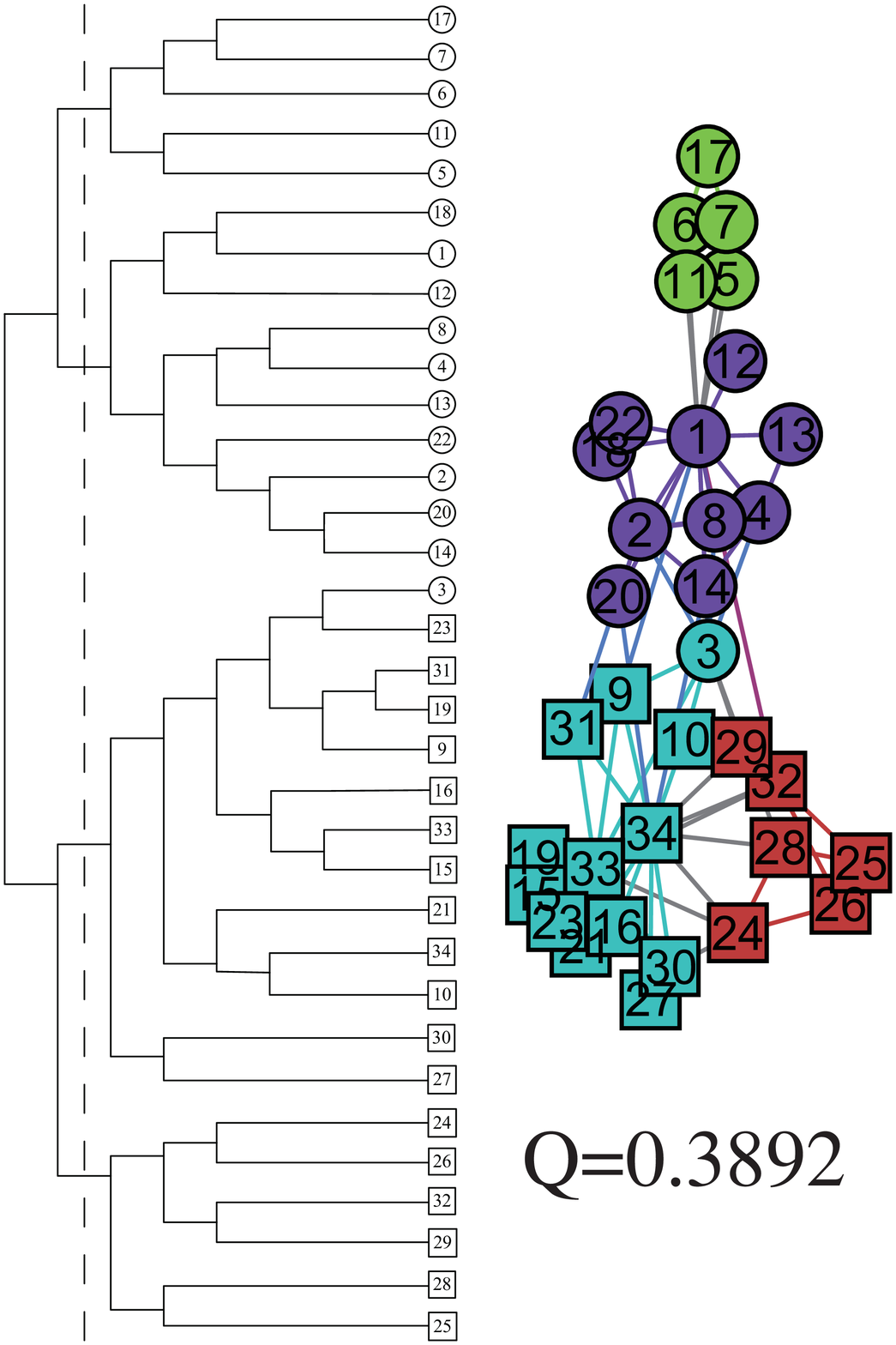}}

 \subfloat[] {\includegraphics[scale=0.3]{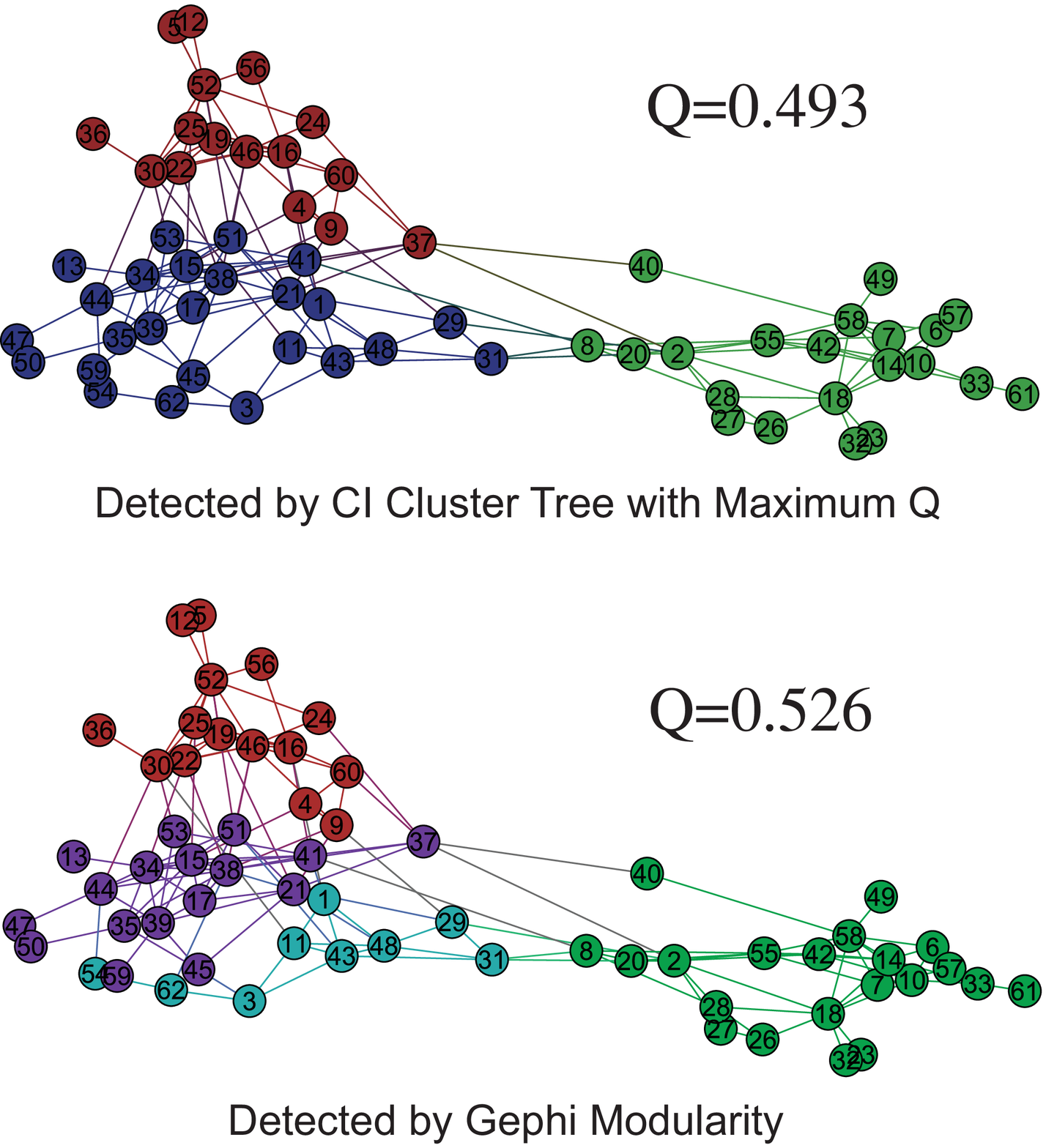}}\label{fg:5}
 \caption{(a) is the cluster tree of Zachary Network by $L^1$ norm CI method with $1/k$ correction. The partition by the dashed has the maximum Q. $\square$ and $\bigcirc$ refers to vertexes of the two divided clubs in history. Different colors marks different clusters of Zachary Network. (b) shows Dolphin Network clustering by $L^1$ norm CI method with 1/k correction. Different colors marks different clusters. We use modularity tool of Gephi to detect clusters on Dolphin Network. We set mode random and resolution 1.}
\end{figure}
For our method is based on numerical optimization, among solutions of each computing, usually, there is a little difference always observed in the overlapping vertexes or between clusters with close relation.\\\indent
It should be pointed out that we cannot conclude our method is worse than that of Gephi(based on \cite{22}) for the reason that $Q$ of ours is less than $Q$ of Gephi. No one can prove $Q$ function is the most appropriate evaluation criterion. In fact, there are different opinions about $Q$\cite{19}. Nevertheless, $Q$ partly has validity which is verified by a lot of real networks. But focusing on the little difference of $Q$s of two methods is meaningless.
\section{Discussion}
Theoretically, we can expand the method to directed graph and weighted graph, for \textquoteleft undirected' or \textquoteleft unweighted\textquoteright~ is not the necessary condition. We just need to replace adjacent matrix with weighted adjacent matrix for weighted graph.
\\\indent
Based on DEI and movement correlation, we can put forward a different method as follow:
\begin{enumerate}
  \item Derive optimal DEI
  \item Extend the feasible region of objective function $Z$ to $R^N$. The solution vector is a list of  coordinates of all vertexes. Randomly select a small part of vertexes and with certain probability, add a small displacement to the optimal DEI coordinates of these vertexes selected (e.g. with uniform distribution in $-0.1\thicksim0.1$). Fix these vertexes and calculate the gradient direction of $Z$ taking the rest vertexes as arguments. Multiply the minus gradient direction with the total displacement and add it to the solution vector.
  \item Repeat step 2. Each time we can get a $N$ dimensional column vector, and they together form a matrix. Calculate all correlation coefficients between any two rows of the matrix, then we can get the correlation matrix $R$.
  \item Rearrange the vertexes which corresponds to the elements of $R$ in the order of optimal DEI.
\end{enumerate}
This method will lesson the time complexity greatly comparing with the method mentioned above. The validity should be tested in future work.
\\\indent
In this paper, we imbed AG in 1-D \textquoteleft line' space. While we can imbed that in high dimensional spaces or some spaces with very different topological structures (e.g. 1-D \textquoteleft ring\textquoteright space \cite{20}). What is optimum structure? This is a question.
Considering the high dimensional \textquoteleft line' space, maybe there is some $M$, in any \textquoteleft line' space whose dimension is higher than $M$, the configurations of the optimal imbedding are the same, which is called faithful imbedding. $M_0$, the infimum of $M$, is able to reflect the complexity of clustering structure(E.g., we can define the quantity $M_0/(N-1))$. Otherwise, we can coarse grain the configurations of faithful imbedding, each grain is a cluster and coarse grained topologies shows the network's skeleton. There are a lot of related questions worth considering and studying.

\bibliography{ppp}
%% \begin{table}
%% \caption{Repeat length of longer allele by age of onset class}
%% \begin{tabular}{@{\vrule height 10.5pt depth4pt  width0pt}lrcccc}
%% table text
%% \end{tabular}
%% \end{table}

%% For two column figures and tables, use the following:

%% \begin{figure*}
%% \caption{Almost Sharp Front}\label{afoto}
%% \end{figure*}

%% \begin{table*}
%% \caption{Repeat length of longer allele by age of onset class}
%% \begin{tabular}{ccc}
%% table text
%% \end{tabular}
%% \end{table*}
%\end{article}
%----------------------------------------------------------------------------------------

\end{document}